\RequirePackage{fix-cm}
\documentclass[twocolumn,epjc3]{svjour3}  
\smartqed  % flush right qed marks, e.g. at end of proof
\RequirePackage{graphicx}
\usepackage{aas_macros}
\usepackage{ulem}
\usepackage{xcolor}
\journalname{Eur. Phys. J. C}
\begin{document}

\title{Radio interferometry applied to the observation of cosmic-ray induced extensive air showers.}
% \subtitle{Do you have a subtitle?\\ If so, write it here}

% \titlerunning{radio }        % if too long for running head

\author{Harm Schoorlemmer \thanksref{e1,addr1,addr3,addr4}
        \and
        Washington R. Carvalho Jr.\thanksref{addr2,addr3,addr4} %etc.
}

\thankstext{e1}{e-mail: hschoorlemmer@science.ru.nl}


\institute{Max-Planck-Institut f\"ur Kernphysik,
 Saupfercheckweg 1, 69117, Heidelberg, Germany \label{addr1}
           \and
           Physics Institute, University of S\~ao Paulo, Rua do Mat\~ao 1371, S\~ao Paulo, Brazil \label{addr2}
           \and 
            IMAPP, Radboud University Nijmegen, Nijmegen, The Netherlands
           \label{addr3}
           \and 
            Nationaal Instituut voor Kernfysica en Hoge Energie Fysica (NIKHEF), Science Park, Amsterdam, The Netherlands
           \label{addr4}
        %   \and
        %   \emph{Present Address:} if needed\label{addr3}
}

\date{Received: date / Accepted: date}

\maketitle

\begin{abstract}
We developed a radio interferometric technique for the observation of extensive air showers initiated by cosmic particles. In this proof-of-principle study we show that properties of extensive air showers can be derived with high accuracy in a straightforward manner. When time synchronisation below $\sim$1\,ns between different receivers can be achieved, direction reconstruction resolution of $< 0.2^\circ$ and resolution on the depth of shower maximum of $<10$\,g/cm$^2$ are obtained over the full parameter range studied, with even higher accuracy for inclined incoming directions. In addition, by applying the developed method to dense arrays of radio antennas, the energy threshold for the radio detection of extensive air showers can be significantly lowered. The proposed method can be incorporated in operational and future cosmic particle observatories and with its high accuracy it has the potential to play a crucial role in unravelling the composition of the ultra-high-energy cosmic-particle flux.
\keywords{Cosmic Rays \and Radio Detection \and Reconstruction algorithms}
\end{abstract}

\section{Introduction}
Interferometry is a method to expose coherence properties in wave phenomena and is applied in many branches of physics. In this study we show how interferometry can be used to derive properties of a relativistic cascade of particles initiated by  an (ultra-) high-energy cosmic particle in the atmosphere, a so-called extensive air shower (EAS).\\ 
By observing properties of EASs, like the primary axis of propagation and the particle density along that axis or at ground level, the energy, incoming direction and composition of the cosmic particles can be derived. Classically, the characteristics of an EAS can be observed by recording the flash of faint fluorescence and/or Cherenkov light emitted during its propagation through the atmosphere and/or by sampling the footprint of the EAS using particle detectors at ground level. However, when an EAS propagates through the atmosphere it also emits radio-waves which can be observed by an array of antennas. For recent reviews on the radio detection of cosmic ray induced air shower see~\cite{ReviewHuege,SCHRODER20171}. From a macroscopic point of view, the emission can be understood as the sum of two major contributions, one arising from the deflection of the secondary electrons and positrons in the geomagnetic field and the other arising from negative charge build-up in the shower front~\cite{2014PhRvD..89e2002A,ALVAREZMUNIZ201429}. Radiation emitted from different regions in the EAS development will add up coherently when the differences in arrival time at the receiver are smaller than a quarter of the wavelength. The power of radiation that is received coherently depends on the location of the observation. As the relativistic moving emitter propagates through the refractive atmosphere it leads to a Cherenkov-cone feature in the radiation pattern where the received radiation is maximally coherent \cite{2011PhRvL.107f1101D,2015APh....65...11N}.\\
Although radio interferometry is widely used in radio astronomy, the properties of signals from radio emission from EAS mandate a different approach. A key difference is that the emitting region is typically not in the far-field where the wavefront can be approximated by a plane. As will be shown, this difference will actually enable three dimensional reconstruction of the EAS properties. 
Another difference is that the signal is impulsive (with a typical duration in the order of nano-seconds), which has the benefit that aliasing features occurring at time differences that are multiples of the wavelength are suppressed in comparison to continuous emitting sources. The final difference is that the polarisation of the negative charge build-up emission depends on the location of the receiver. We choose not to address this issue for now as it is typically the sub-dominant contribution to the overall emission, but we recognise that a proper treatment could be implemented as an improvement to the proof-of-principle study presented here.

\section{Methodology}
In our method, waveforms $S_{i}(t)$ measured with receivers at different locations $i =(1,..,n)$ are added in the following way 
\begin{equation}
S_{j}(t) = \sum_{i}^nS_{i}(t-\Delta_{i,j}),
\label{eq:super}
\end{equation}
with $\Delta_{i,j}$ the light-propagation time from location $j$ to $i$. $S_j(t)$ present the coherent waveform at location $j$ and is the key quantity in the interferometric method. %W3:Left as it was: S_j IS the (one) key constant
Since we are interested in impulsive signals, we need the actual travel time of the light $\Delta_{i,j}$, rather than the geometrical phase difference which is sufficient for continuously emitting source in the far-field and commonly used in radio astronomy. By evaluating the power in $S_{j}$ at different locations $j$ a three-dimensional measure of the coherence is obtained. The light travel time $\Delta_{i,j}$ from each antenna location $i$ to location $j$ is approximated by
\begin{equation}
\Delta_{i,j} \approx \frac{d_{i,j}\bar{n}_{i,j}}{c},
\end{equation}
with $c$ the speed of light in vacuum, $d_{i,j}$ the distance between location $i$ and $j$, and $\bar{n}$ the average atmospheric refractive index along the path from $i$ to $j$, 
assuming that the light travels in straight paths between $i$ and $j$. This is a valid approximation for the calculations of radio emission from air showers (see appendix A of \cite{ALVAREZMUNIZ201531}). The atmospheric refractive index can be parametrised as an exponential function of altitude $z$, $n = 1 + a e^{-bz}$, and for simplicity in the remainder of the paper all the calculations are performed using $a=325\times10^{-6}$ and $b = 0.1218$\,km$^{-1}$. However, it has been tested that the method is robust against (small) deviations from the assumed refractive index model that can occur during regular atmospheric conditions.\\ 
\begin{figure}[h]
\includegraphics[width=0.99\columnwidth]{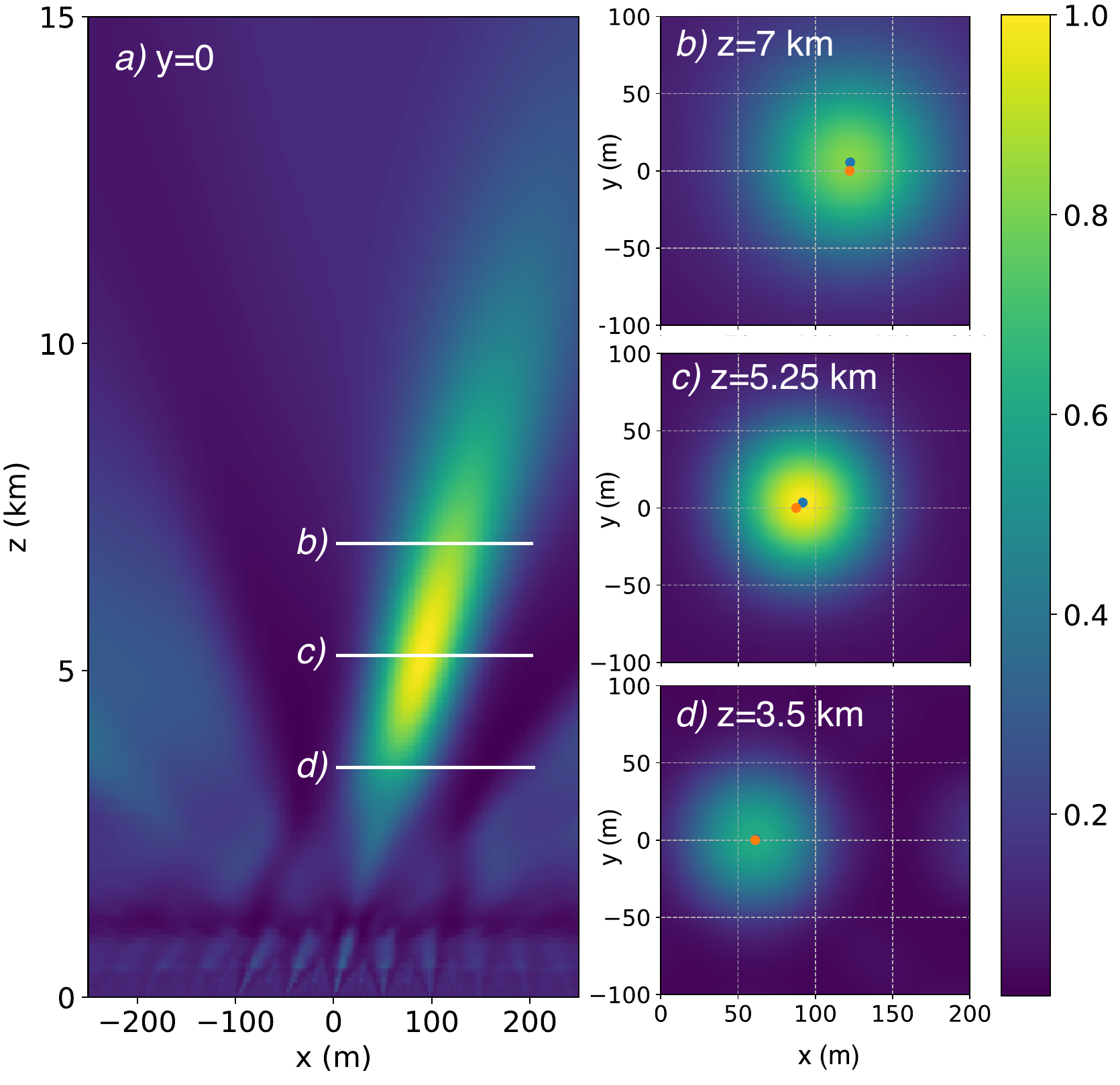}\caption{\label{fig:example_event} The Radio interferometric technique applied to a simulation of a cosmic-ray induced air shower initiated by a $10^{17}$\,eV proton with an incoming direction one degree from zenith.  Panel \textit{a)} shows the %W:added same plane
normalised power of $S_{j}(t)$ in Eq. \ref{eq:super} mapped onto the vertical plane that contains the shower axis (chosen to be at y=0), while \textit{b}), \textit{c)} and \textit{d)} show horizontal planes at different heights.  As receivers (measuring each $S_{i}(t)$ in Eq. \ref{eq:super}) we used an array of 25 antennas on a grid with equal spacing of 100\,m.  In \textit{b)}, \textit{c)} and \textit{d)} the orange dot marks the location of the true shower axis, while the blue dot marks the maximum in the map.}
\end{figure}
We note here that interferometric approaches have been tried  before, especially the early work by \cite{NIGL} comes very close to the approach presented here. In \cite{NIGL}, and also \cite{WAVEFRONT,2021EPJC...81..176A}, the arrival times of the signals are approximated to follow a parametrised "radio wavefront". Missing in these approaches is an exact calculation of the travel time from emitter to receiver that takes into account the varying atmospheric refractive index, which is a crucial part of the method described in this work.

\section{Application to air shower reconstruction}
To study the spatial distribution of $S_j$, we used the ZHAireS simulation code \cite{ZHAireS} to calculate the radio emission from air showers simulated by the Aires package. 
The ZHAireS calculations provides us with a time-dependent electric field vector at a specified antenna position. To obtain signal traces $S_i(t)$, a  30--80\,MHz bandpass filter has been applied to mimic roughly the frequency response of currently operational arrays like LOFAR~\cite{LOFAR},  AERA~\cite{2014PhRvD..89e2002A}, Tunka-REX~\cite{TUNKA-REX}, and OVRO-LWA \cite{OVRO-LWA}. 
For simplicity, we decided to evaluate the component  of the electric-field in the direction perpendicular to the geomagnetic field orientation and to the direction of propagation of the particle shower. This corresponds to the axis of polarisation of the geomagnetic emission, which is the dominant contribution 
for most EASs.\\
In Figure \ref{fig:example_event}, we show the mapping of the power of $S_j$ integrated in the frequency band in several planes for a simulated 10$^{17}$\,eV proton entering the atmosphere one degree from zenith (we normalised the map to have maximum at 1). The mapping in the $xz$-plane (panel $a$) shows clearly the coherent power being mapped near the shower axis and it reaches a maximum along the shower-axis. The horizontal slices (panels $b$, $c$, $d$) show that the location of maximum power in these maps only deviate a few metres from the true position of the shower axis.\\
\begin{figure*}[ht]
\includegraphics[width=0.32\textwidth]{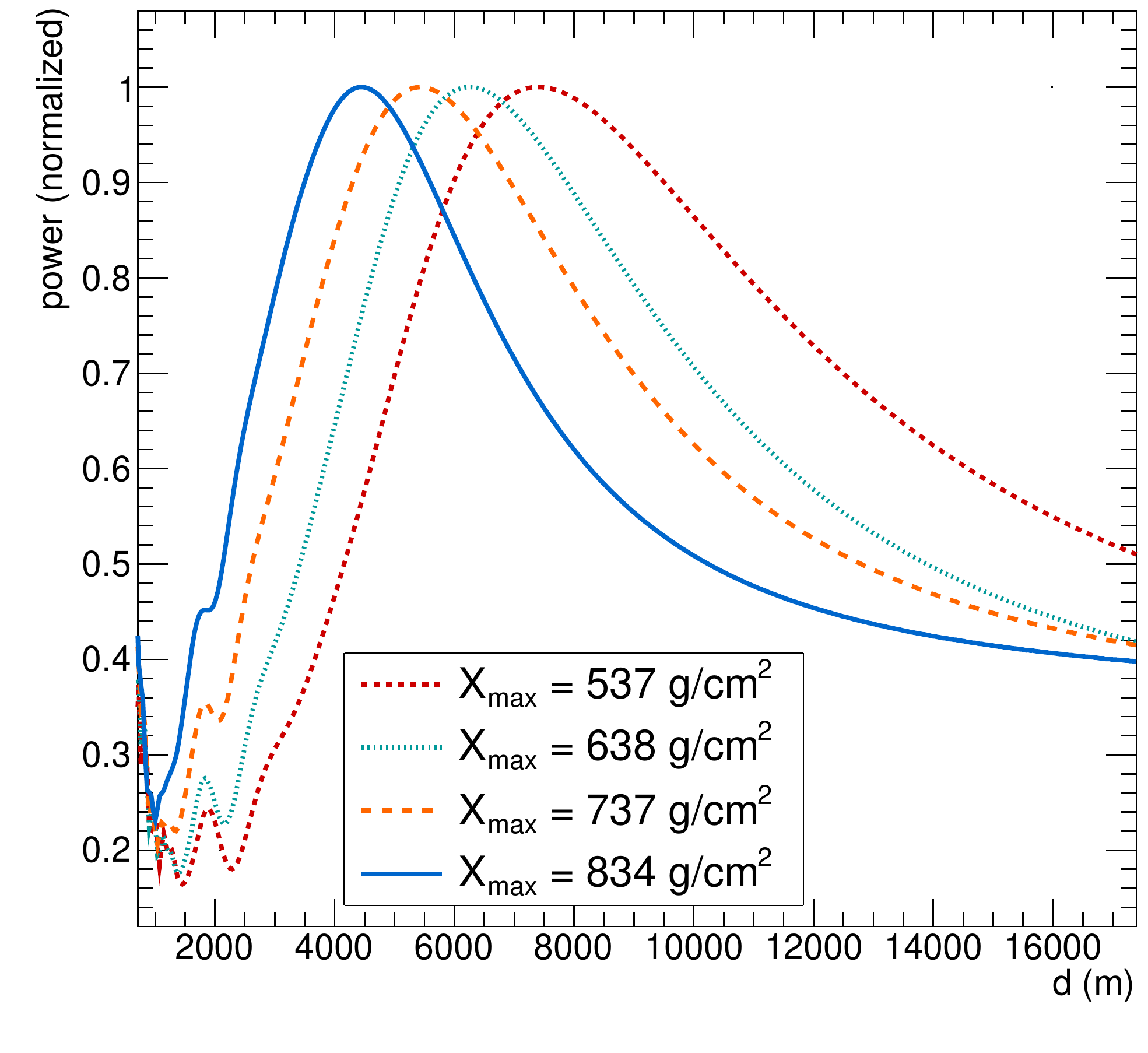}
\includegraphics[width=0.326\textwidth]{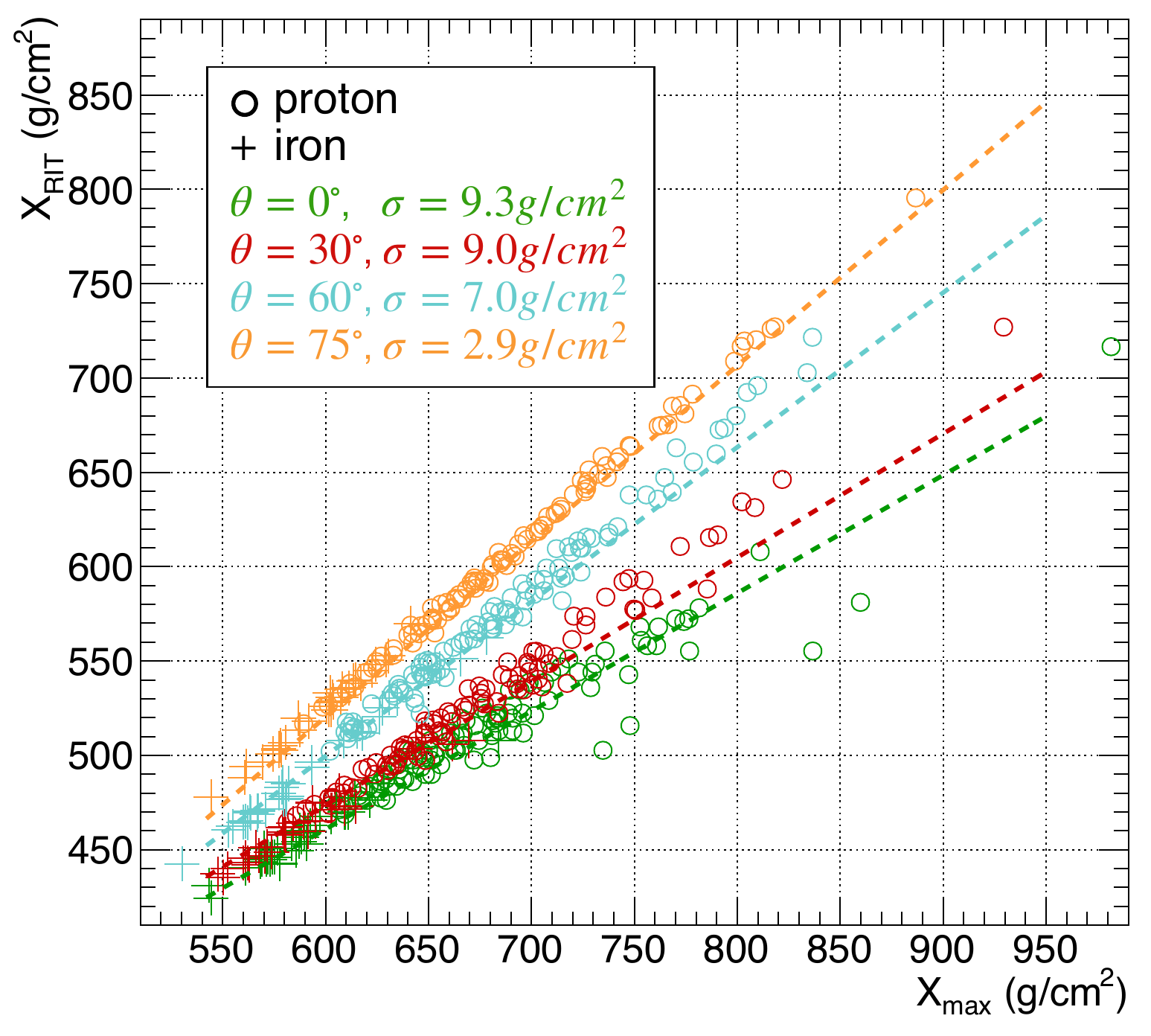}
\includegraphics[width=0.326\textwidth]{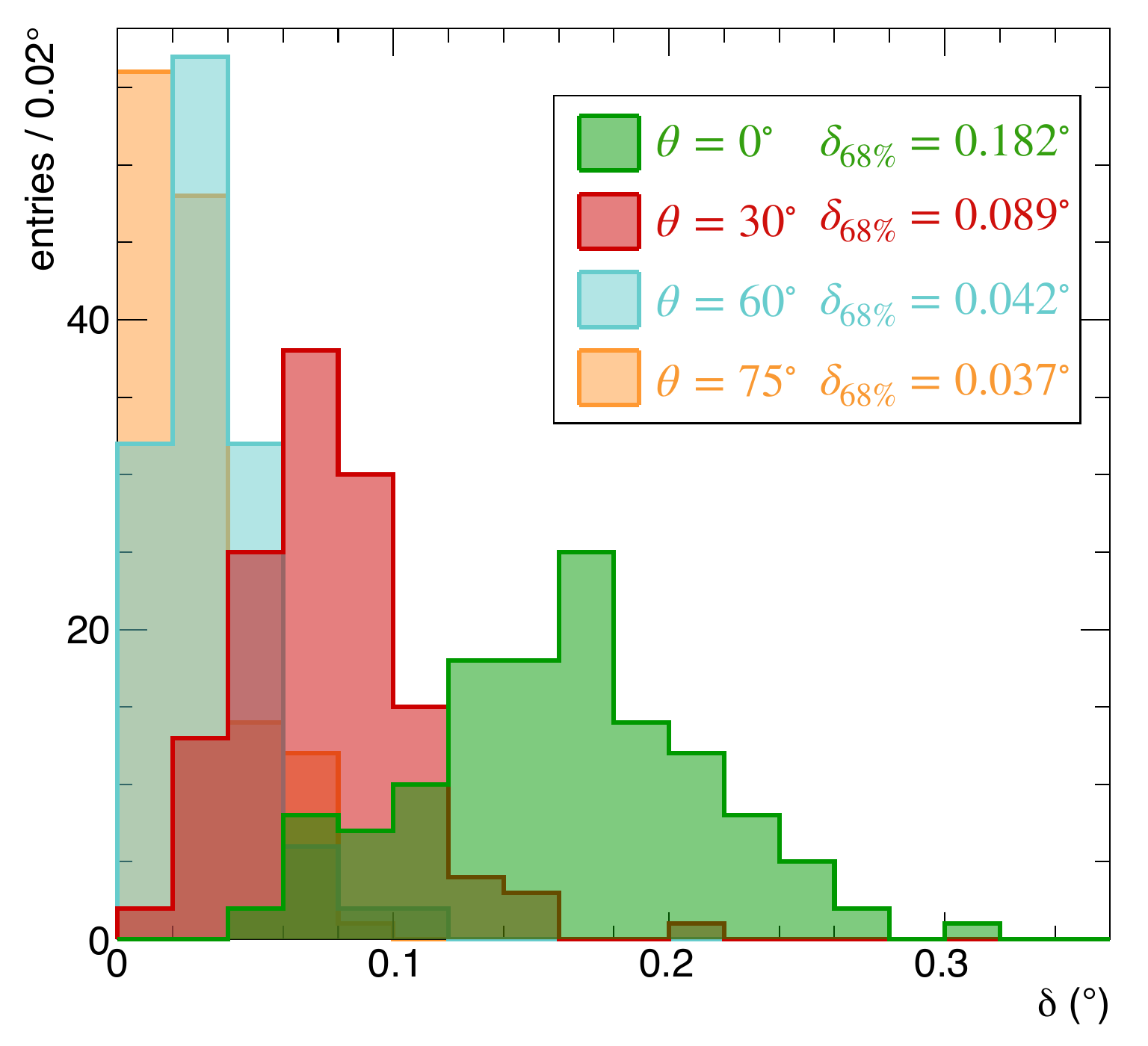}
\caption{Air shower development traced by the radio-interferometric-technique (RIT). Left panel, examples of the power evaluated at distance $d$ measured from the ground  along the shower axis. Four simulated proton induced air showers with a zenith angle of $\theta=30^\circ$ are shown.
Middle panel, correlation between $X_{\textrm{max}}$ and the slant-depth $X_{\textrm{RIT}}$ of the on-axis maximum of the power obtained by the reconstruction for four simulation sets at different zenith angles. Each set consists of 30 iron and 100 proton induced air showers. Right panel, angular deviation $\delta$ between reconstructed and true direction for the same four sets of simulations as in the middle panel.}
\label{fig:longitudinal}
\end{figure*}
We are going to use the observation that the power mapping strongly correlates with the air shower axis to reconstruct air shower properties. We applied the following reconstruction approach to identify an axis in the  power mapped space:
\begin{enumerate}
\item A horizontal plane is defined at an altitude corresponding to the average depth of shower maximum for a reference energy of the primary particle. In this plane an iterative search  for the position of the maximum power is conducted. In each iteration the resolution is adjusted to zoom in on the "hottest"  pixel.
\item  Based on the location of the maximum in step 1, multiple horizontal planes (like in figure 2) are defined and in each plane we search again for the location of the maximum. We fit a track to the locations of the set of maxima in the horizontal planes, what gives the first estimate of the parameters of the shower axis:  the impact point at ground level $x_0,y_0,z=0$ and an azimuth $\phi$ and zenith $\theta$ angle.
\item In a final step, the estimated parameters from step 2 are used as starting point to maximise the integrated power along a track of predefined length by varying the four axis parameters.
\end{enumerate}
We calculated with ZHAireS the radio emission of sets of proton and iron induced air showers at different zenith angles. A rectangular array of antennas was used in which we increased the grid-spacing of the grid unit as a function of the zenith angle of the shower (See Table \ref{tab:1}). In this way, the number of antennas in the radio-footprint on the ground was kept roughly constant (typically  between 25-40) in the different zenith angle sets. 
\begin{table}[h]
\caption{Array properties.}
\label{tab:1}       % Give a unique label
\begin{tabular}{cc}
\hline\noalign{\smallskip}
zenith angle & Antenna density (km$^{-2}$)\\
\noalign{\smallskip}\hline\noalign{\smallskip}
0$^\circ$ & 400  \\
30$^\circ$ & 204 \\
60$^\circ$ & 22  \\
75$^\circ$ & 2.7 \\
\noalign{\smallskip}\hline
\end{tabular}
\end{table}

We randomised the impact point of the shower-axis uniformly within the central rectangular grid unit. The reconstruction approach has been applied to this simulated set of air showers and the main results are summarised in Figure \ref{fig:longitudinal}.\\
In the right panel of Figure \ref{fig:longitudinal} the direction of the reconstructed axis is compared to the location of the true shower axis. The obtainable angular resolution is indicated by the 68\% containment level $\delta_{68\%}$. For both the direction and $X_{\textrm{max}}$ reconstruction we observed that the accuracy improves with increasing zenith angle (and increasing depth of shower maximum), i.e. when the shower is physically closer to the receivers the method works poorer. This is not yet fully understood and investigating this trade-off in more detail will be part of a follow-up study. One plausible reason might be that coherence in the received radiation is not achieved over large scales of the emitting air shower when it is too nearby.\\ 
The result of evaluating the power along the reconstructed axis for four simulated air showers that reach shower maximum at depths that increase in steps of roughly  100\,g/cm$^2$ is shown in the left panel in Figure \ref{fig:longitudinal}. The distribution of the observed power along the reconstructed axis depends clearly on the longitudinal development of the air shower. In the middle panel of Figure \ref{fig:longitudinal} the comparison is shown between the depth $X_{\textrm{RIT}}$ of the maximum power along the reconstructed axis (now expressed as slant-depth in units of g/cm$^2$ ) and the true depth $X_{\textrm{max}}$ where the air showers reach their maximum electron density. $X_{\textrm{max}}$ is one of the key observables in characterising the composition of the cosmic particle flux.
The relation between $X_{\textrm{RIT}}$ and $X_{\textrm{max}}$ at a specific angle can be described by a simple first order polynomial. We used this fit to estimate an $X_{\textrm{max}}$ value for each shower. This estimate was then compared to the true shower maximum to obtain a distribution of the deviations. We then took the RMS of this distribution as the scatter $\sigma$, which represents a measure of the resolution that is achievable by the method. The obtained overall resolution is encouraging, with $\sigma<10$\,g\,cm$^{-2}$ and decreasing significantly with zenith angle. For comparison, the resolution of the fluorescence detection technique by the Pierre Auger Observatory \cite{2014PhRvD..90l2005A} ranges from 25\,g\,cm$^{-2}$ at 10$^{17.8}$\,eV to 15\,g\,cm$^{-2}$ towards the highest energies ($>10^{19}$\,eV), while the Low Frequency Array (LOFAR), using radio measurements and a reconstruction method based on full Monte-Carlo simulations, reported a mean uncertainty of 16 \,g\,cm$^{-2}$ in the energy range from $10^{17}$ to $10^{17.5}$\,eV \cite{LOFARNATURE}.\\
\subsection{Adding background and timing uncertainties.}
\begin{figure}[h]
\includegraphics[width=0.99\columnwidth]{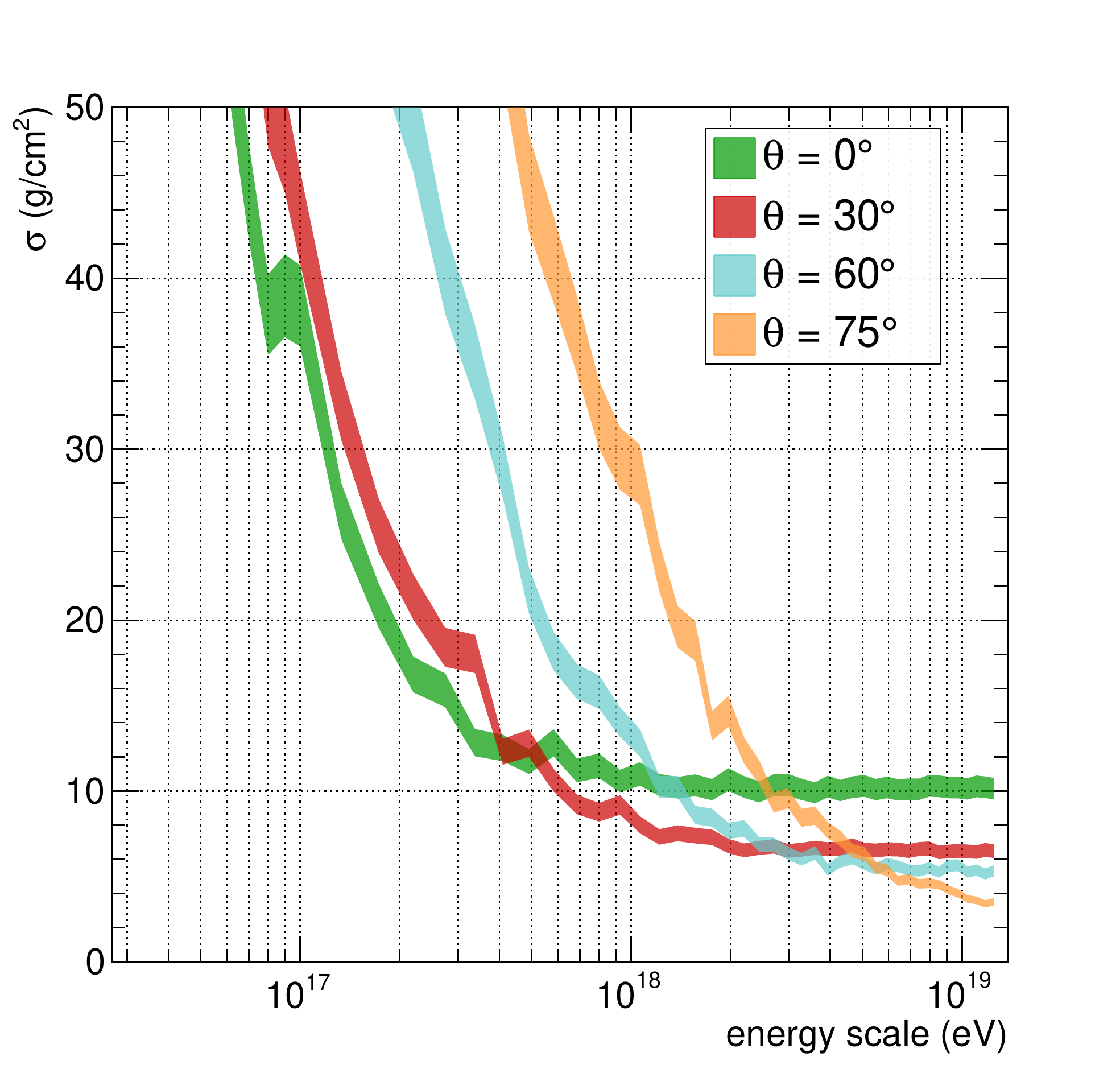}
\includegraphics[width=0.99\columnwidth]{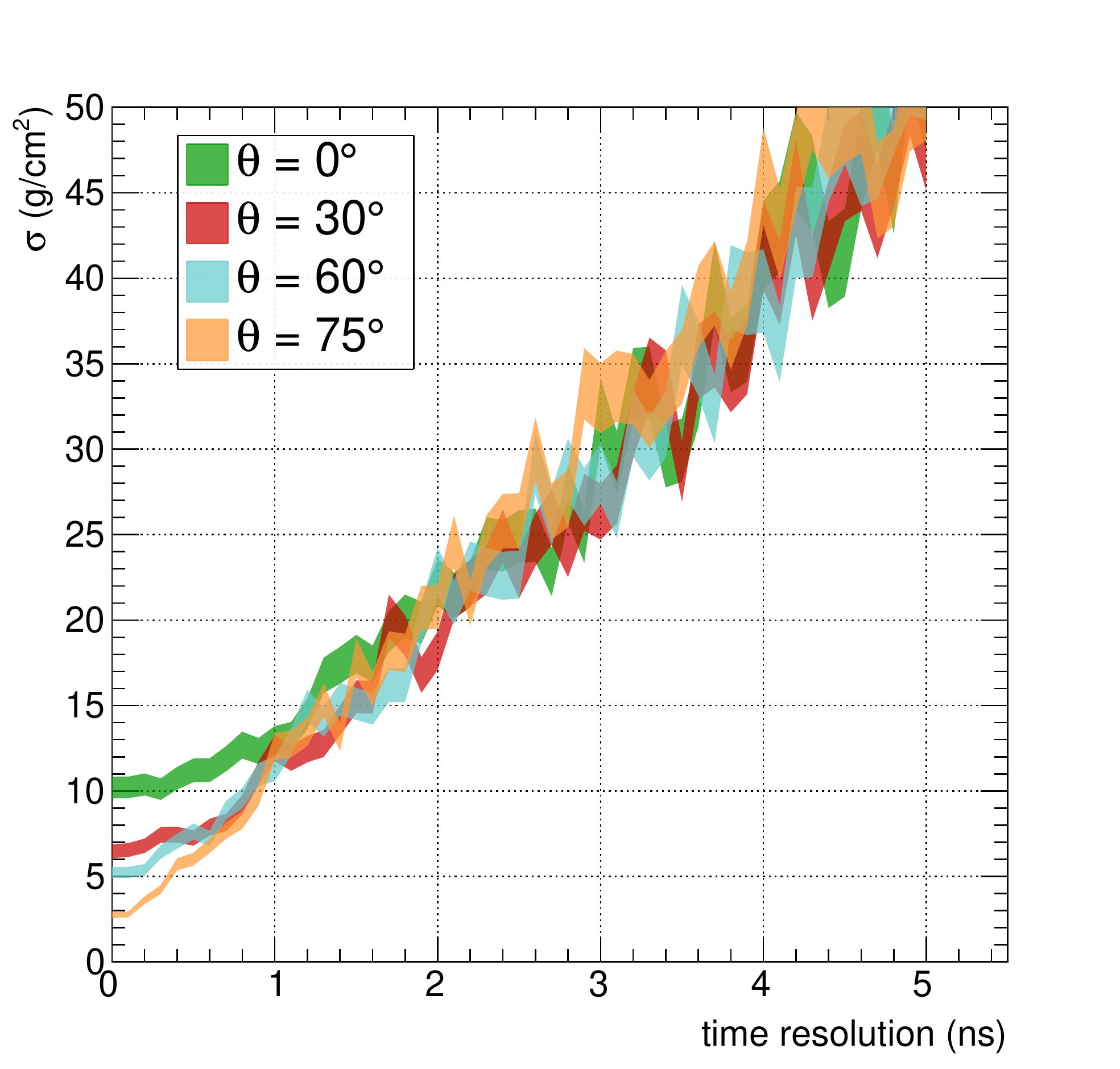}
\caption{\label{fig:xmaxres} Top: Resolution on $X_{\textrm{max}}$ ($\sigma$) as a function of the energy scale of the primary particle for the four zenith angle sets and their corresponding antenna densities from Table \ref{tab:1}. A typical background noise level in the frequency band from 30-80 MHz is considered. Bottom: Resolution on $X_{\textrm{max}}$ ($\sigma$) as a function of receiver-to-receiver time synchronisation resolution (Gaussian width) for individual receivers. No background noise was added.}
\end{figure}
The results presented in Figure \ref{fig:longitudinal} were obtained in the absence of any background radio signals and should therefore be regarded as the case where the influence of background is negligible. However, especially in the frequency range between 30--80\,MHz the Milky Way is providing an omnipresent background contribution, which results in a typical $\sim25\,\mu$V/m (RMS) noise level in this frequency band. In Figure \ref{fig:xmaxres} (top) we added this noise level to the simulations and reconstructed the location of $X_{\textrm{RIT}}$ and estimated the resulting resolution $X_{\textrm{max}}$. The strength of the electric field of the radio signal scales linearly with the energy of the primary particle, we apply this scaling to the reference energy of $10^{17}$\,eV to mimic other energies. Here we warn the reader that the results in Figure \ref{fig:xmaxres} are meant as to illustrate to first order the effect. In general the expected performance will depend on the antenna density (given Table \ref{tab:1}, also see section \ref{sec:thres}), the location of the observatory, the event geometry, and observing conditions. We see that the performance converges to the most accurate resolution with increasing energy, the energy at which this happens depends on the configuration of the simulations that were used in this paper. In addition, the presence of background mandates a search strategy for the signal in the summed waveform ($S_j$). Here we have chosen a pragmatic search window of 200\,ns based on the event geometry, which in a real experiment can be tested from either from the radio signal or an external trigger. For implementation in specific experiments this search window should be optimised and can either improve or degrade the obtainable resolution at given energy.\\  

Since this method both utilises the amplitude and timing information in the signal, it is expected that synchronisation error from station to station will deteriorate the performance. A simple constrain on synchronisation comes from the coherence condition, which implies that the timing uncertainty $\Delta_t$ and position uncertainty $\Delta_x$ of each receiver needs to be known to an accuracy of smaller than quarter of the wavelength. For a signal at 50\,MHz this yields $\sqrt{\Delta_t^2 + (\Delta_x/c)^2 }< 5$\,ns. We verified that synchronisation uncertainties beyond this limit will destroy the coherence mapping. However, timing jitter on individual stations already impacts the achievable resolution on reconstructed air shower parameters at smaller values. We illustrate this in Figure 3 (bottom) where we added Gaussian distributed random timing jitter to each simulation and obtained the resolution on $X_{\textrm{max}}$. One can observe above 1\,ns timing jitter, a linear degradation of the resolution of shower maximum. We note here that this result is valid for the specific simulation setup used here, but can be taken as an indication of the right scale. The timing accuracy needed to get competitive results (below $~2$\,ns) is beyond the achievable accuracy of individual GPS receivers. Therefore, this method either needs cabled setups and/or the use of external synchronisation signals.  We note here that made a pragmatic choice to use a Gaussian distribution to sample the time uncertainty, which might not reflect realistic time uncertainty models.\\
Dedicated studies of simulated showers including full instrumental response and array configurations are needed to estimate the performance of this method in the different operational and planned experiments. However, the results from the relatively simple reconstruction approach given here yield a promising starting point for more elaborate approaches including more realistic instrument modelling. An example of such a study can already be found in \cite{Schl_ter_2021}.\\
In the study presented here, we used regular grid arrays. However, for highly irregular grids, or very sparse arrays, the simple linear relation for a given zenith angle will no longer hold. Like in traditional radio-astronomy, you need to have enough baselines to sample the relevant dimensions of the emission region sufficiently. However, having prior knowledge of the expected emission region (which we have through MC simulations), can be used in a fitting approach to partially compensate for the lack of baselines in some observations. This idea will be explored in future work for specific implementation of radio antenna arrays.\\

\begin{figure*}[!ht]
\includegraphics[width=0.32\textwidth]{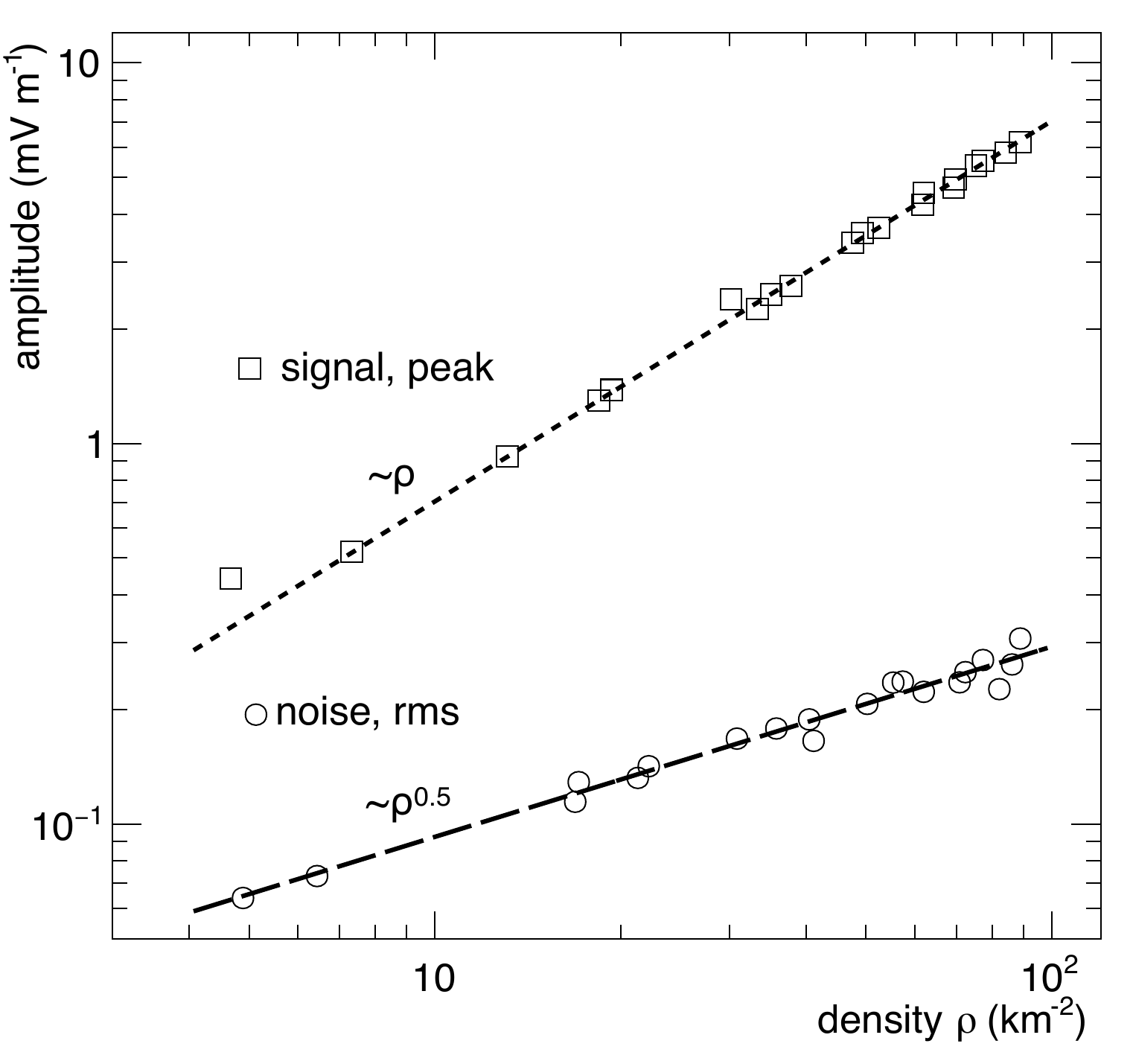}
\includegraphics[width=0.328\textwidth]{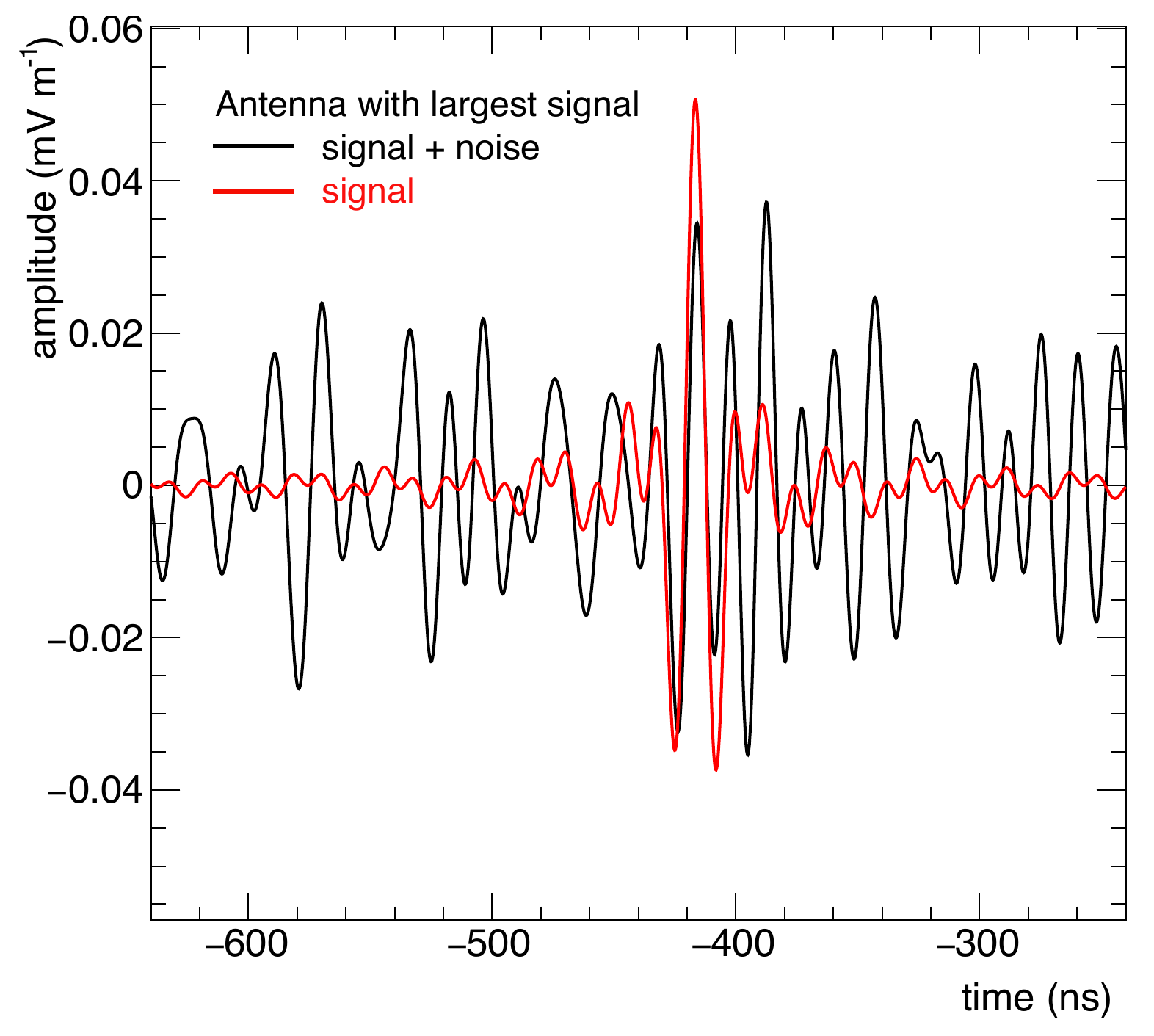}
\includegraphics[width=0.32\textwidth]{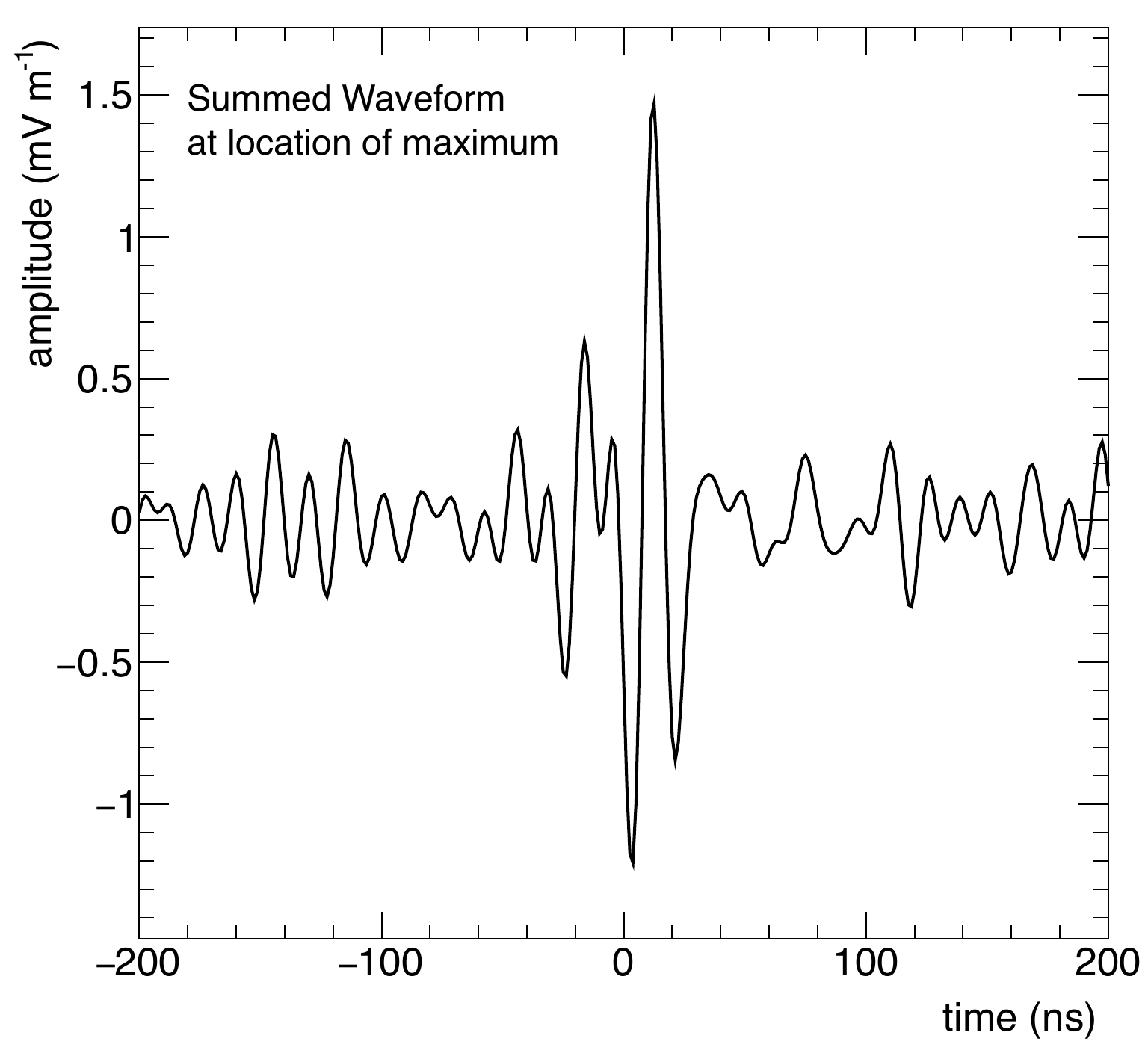}
\caption{Properties of the summed waveform $S_{j,\textrm{max}}$ with the maximum power along the shower axis for  a 10$^{17}$\,eV proton induced air shower 60$^\circ$ from zenith. At the location of each antenna  a waveform with white-noise is simulated with a root mean square of 14\,$\mu$V/m.  Left panel, dependency of the amplitude of $S_j$ at $X_{\textrm{RIT}}$ on the antenna density of the array.  Middle panel, the waveform of the location on the ground that received the maximum signal from the air shower. Right panel, the summed waveform $S_{j,\textrm{max}}$ for an  array density of 22 antennas per km$^2$.}
\label{fig:interometry}
\end{figure*}
This radio interferometric technique provides a straight-forward way of reconstructing EAS parameters, with the only assumption being the atmospheric refractive index model. As shown, the mapping of the power in $S_j$ gives access to shower parameters like the primary axis and longitudinal development parameters. However, this interferometric method also has its limitations and artefacts. For example, one artefact shows up in the left panel of Figure \ref{fig:longitudinal}, where there is a long tail in the power of $S_j$ that remains even in the region of large distances along the shower axis, where the number of particles from air shower development vanishes. This is caused by the fact that the gradient of arrival times reduces to zero for antennas in the array for large distances $d$ along the shower axis. Therefore there is some remainder power, which  corresponds to the power of $S_j$ under a plane wave approximation.\\ 
The radio interferometric technique has the appearance of being a tomographic method as it can be applied in three dimensions. However, one should use this term with some caution, since  the mapping of the power in $S_j(t)$ is not a direct representation of the emitted radiation, the spatial data points are correlated over scales set by coherence condition for a given wavelength. Applying the method at higher frequencies than used in this study will shorten the coherence length and therefore will give a more direct mapping of the emitting regions of the EAS. In addition, it might be possible to expand the method with deconvolution techniques in order to map the emission directly.

\section{Application to air shower detection threshold \label{sec:thres}}
One important aspect of interferometric methods is that signal-to-noise ratio scales as the square-root of the number of receiving elements. This means that the detection threshold of the radio-detection technique can be lowered by increasing the antenna density in the array accordingly. This behaviour is illustrated in Figure \ref{fig:interometry} in which we vary the antenna array density for a 60$^\circ$ simulated air shower. At each receiver, we simulate a random white noise realisation resulting in an average root mean square of 14\,$\mu$V/m. 
In the left panel of Figure \ref{fig:interometry} the summed waveform $S_{j,\textrm{max}}$ at $X_{\textrm{RIT}}$ is evaluated as a function of receiver density in the array.  We show two cases, once simulating noise and once only simulating signal. As expected for the coherent signal, the peak amplitude of  $S_{j,\textrm{max}}$ scales linearly with the density of the receivers while the incoherent noise level increases as the square-root of the receiver density. As an example of how this could lower the energy threshold for estimating air shower parameters, we show in the middle panel the waveform of the antenna that receives the strongest signal in the array. 
As is visible, when the noise is added to the signal (black line)  it is hard to identify the presence of any signal at all. However, a clear signal can be seen in the coherent sum, when all the antennas in the array are combined to obtain $S_{j,\textrm{max}}$, as shown in the right panel. There are existing high-density arrays that are capable of measuring EAS, like the LOFAR detector where this method can be directly tested. New facilities, like the square kilometre array,  might have a density of ~60000 antennas per km$^2$\cite{SKA}. For such arrays, this analysis technique will enable the study of air shower parameters over a significantly larger range of energies, provided an external trigger is obtained from a simple array of particle  detectors. It is unlikely that this method can be used to trigger on radio signal from cosmic ray events as unbiased scanning of three dimensional space requires intensive computing. However, this method might be used to identify the location of noise source offline. Events from stationary noise sources will result in fixed time offsets between receivers, which could be used to veto events these events in an online trigger (or offline analysis). 
 
\section{Discussion}
Several upcoming experiments are focusing on radio observation of inclined air showers, where the interferometric method is showing its best performance. The Pierre Auger Observatory is currently being upgraded which includes a radio-antenna on each water-Cherenkov 
tank of the surface detector in order to observe the radio emission of inclined events~\cite{2018JCAP...10..026A}. The determination of shower maximum with the method developed here will enable accurate hybrid measurements of inclined air showers to be used in composition studies. Another project is the Giant Radio Array for Neutrino Detection \cite{2020SCPMA..6319501A}, where the radio interferometric method might both help with air shower reconstruction and with the identification of air showers among other impulsive events. Since both these arrays are deployed over large areas, the challenge will be to obtain synchronisation between the individual stations accurately enough to satisfy the coherence conditions. This might for example be achieved with the use of external continuous narrow-band transmitters \cite{LOFAR_BEACON,FRANK_BEACON}. In addition, the spacing of the antenna grid for these large arrays is in the ~km scale, resulting in the order of 10 antennas sampling the radio footprint. We tested our method on a square grid with 1\,km spacing at a zenith angle of 75$^\circ$ which resulted in some degradation in the resolution on $X_{\textrm{max}}$ ($\sigma \approx 10$\,g\,cm$^{-2}$), which is still pretty compelling for these very low-density arrays. The reason for this degradation is that below a certain antenna density, or very asymmetric sampling of the radio footprint, not the full information of the shower development is sampled with the baselines contained in the array. This will cause distortions in the three dimensional mapping of the coherent power, hence the straightforward linear correlation with air shower parameters will be reduced. However, there might be the possibility to recover loss of accuracy by adding additional known information on the radio emission into a fitting algorithm.\\
In this study, we explored the relation of the radio interferometric mapping with the direction and the shower maximum from simulations. However the map in is a new observable that might be explored in more detail in itself,  and/or in concert with other measurements, to obtain more information about the primary cosmic particle. For example, in the upgrade at the Pierre Auger Observatory this method might be complemented with the measurement of muon density at ground level. Measuring the longitudinal shower development using radio has almost a continuous duty-cycle and therefore the upgrade of the Pierre Auger Observatory, but more notable GRAND, will help increasing  statistics on the composition of the highest part of the cosmic-ray energy spectrum. At lower energy ($10^{16}-10^{18}$\,eV) it is the high-density of antennas in radio telescopes that might provide very accurate composition measurements in the energy regime where the cosmic-ray flux is expected to transition from mainly galactic to extra-galactic origin.\\ 
We would like to remark that the presented method here is generic and is applicable to a wider range of observations. A similar method has already been applied to the observation of lightning \cite{LOFAR_LIGHTNING}, revealing substructures with unprecedented resolution. Another obvious candidate to apply the radio interferometric technique are the in-ice radio antenna arrays deployed for neutrino observations \cite{SCHRODER20171}. In this case, the method might need to be complemented by ray-tracing algorithms in order to determine the light travel time to sufficient accuracy.\\
To conclude, the method presented here provides a whole new perspective on observations of extensive air showers over a wide energy range, which could lead to significant improvement on the accuracy of the measurements of cosmic particle properties. Therefore this method can play a crucial role in understanding the origin and propagation of these cosmic particles in the near future.   

\begin{acknowledgements}
We would like to thank Jaime Alvarez-Mu\~niz, Stephanie Ann Wissel and Enrique Zas for their insightful comments. W.R.C. thanks grant 2015/15735-1, S\~ao Paulo Research Foundation (FAPESP). 
\end{acknowledgements}
\bibliographystyle{spphys} 
\bibliography{references.bib}% Produces the bibliography via BibTeX.

\end{document}